\documentclass[11pt,a4paper]{article}

\pdfoutput=1

\usepackage{graphicx}
\usepackage{comment}
\usepackage{epstopdf}
\usepackage{jcappub}

\usepackage{amsmath,amssymb,amsbsy,exscale}
\usepackage{graphicx}
\usepackage{epsfig}
\usepackage{multicol}
\usepackage{mathrsfs}
\usepackage{mathtools}
\usepackage{blindtext}
 \usepackage{fancyhdr}
\usepackage{enumerate}

\usepackage{dsfont}
\usepackage{appendix}

\usepackage{parskip}

\usepackage{xspace}
\usepackage{braket}
\usepackage{array}
\usepackage{feynmp}

\usepackage{graphicx,rotating,colortbl}

\usepackage{pinlabel}       

\usepackage{amsfonts}
\usepackage{slashed}

\usepackage{verbatim}
\usepackage{doi}
\usepackage{url}
\usepackage{array}

\usepackage{float}
\usepackage[labelfont=bf]{caption}
\usepackage{subcaption}
\usepackage{parskip}

\usepackage{hhline}

\usepackage{sectsty}

\usepackage{mdframed}
\usepackage{titletoc}

\newcommand{\gappeq}{{\rlap{{\raise}.5ex\text{\ensuremath{>}}}{{\lower}.5ex\text{\ensuremath{\sim}}}}}
\newcommand{\lappeq}{{\rlap{{\raise}.5ex\text{\ensuremath{<}}}{{\lower}.5ex\text{\ensuremath{\sim}}}}}

\newcommand{\I}{\tmtextrm{1{\kern}-.24em l}}

\definecolor{nicered}{rgb}{0.7,0.1,0.1}
\definecolor{nicegreen}{rgb}{0.1,0.5,0.1}

\newcommand{\newc}{\newcommand}
\newc{\D}{\partial}
\newc{\som}{\sin\omega}
\newc{\com}{\cos\omega}
\newc{\sth}{\sin\theta}
\newc{\cth}{\cos\theta}
\newc{\stom}{\sin^2\omega}
\newc{\ctom}{\cos^2\omega}
\newc{\stth}{\sin^2\theta}
\newc{\ctth}{\cos^2\theta}
\newc{\ie}{{\it i.e.} }
\newc{\eg}{{\it e.g.} }
\newc{\etc}{{\it etc.} }
\newc{\etal}{{\it et al.}}

\newcommand{\dd}{\text{d}}

\newcommand{\lapproxeq}{\lower .7ex\hbox{$\;\stackrel{\textstyle
<}{\sim}\;$}}
\newcommand{\gapproxeq}{\lower .7ex\hbox{$\;\stackrel{\textstyle
>}{\sim}\;$}}
\newcommand{\stackdown}[2]{\lower 1.4ex\hbox{$\;\stackrel{\textstyle{#1}}
{\scriptstyle{#2}}\;$}}

\newcommand{\epsh}{\epsilon_{\rm H}}
\newcommand{\epsv}{\epsilon_{\rm V}}
\newcommand{\eps}{\epsilon}

\newcommand{\phib}{\bar{\phi}}
\newcommand{\etah}{\eta_{\rm H}}
\newcommand{\etav}{\eta_{\rm V}}

\newcommand{\vb}{\bar{V}}
\newcommand{\mpl}{M_{\rm Pl}}

\title{Constant-Roll (Quasi-)Linear Inflation}

\author[a]{A. Karam,}

\author[b]{L. Marzola,}

\author[a]{T. Pappas,}

\author[b]{A. Racioppi,}

\author[a]{K. Tamvakis}
 
\affiliation[a]{Department of Physics, University of Ioannina, GR–45110 Ioannina, Greece}

\affiliation[b]{National Institute of Chemical Physics and Biophysics, R\"avala 10, 10143 Tallinn, Estonia}

\emailAdd{alkaram@cc.uoi.gr}

\emailAdd{luca.marzola@cern.ch}

\emailAdd{thpap@cc.uoi.gr}

\emailAdd{antonio.racioppi@kbfi.ee}

\emailAdd{tamvakis@uoi.gr}

\abstract{In constant-roll inflation, the scalar field that drives the accelerated expansion of the Universe is rolling down its potential at a constant rate. Within this framework, we highlight the relations between the Hubble slow-roll parameters and the potential ones, studying in detail the case of a single-field Coleman-Weinberg model characterised by a non-minimal coupling of the inflaton to gravity. With respect to the exact constant-roll predictions, we find that assuming an approximate slow-roll behaviour yields a difference of $\Delta r = 0.001$ in the tensor-to-scalar ratio prediction. Such a discrepancy is in principle testable by future satellite missions. As for the scalar spectral index $n_s$, we find that the existing 2-$\sigma$ bound constrains the value of the non-minimal coupling to $\xi_\phi \sim 0.29-0.31$ in the model under consideration.}

\keywords{Inflation, non-minimal coupling, loop corrections, constant roll}

\begin{document}

\maketitle

\section{Introduction}

According to the theory of cosmic inflation~\cite{Starobinsky:1980te,Guth:1980zm,Linde:1981mu,Albrecht:1982wi}, our Universe underwent a period of exponential expansion during the first instants after its birth. The reasons for considering such a possibility stem from several fine-tuning problems of traditional hot big bang cosmology, namely the horizon and flatness problems, which find a natural solution in the proposed accelerated growth. A fast expansion, in fact, depletes any initial curvature contribution to the energy balance of the Universe and allows for arbitrarily wide particle horizons, resulting in the remarkable temperature uniformity observed at large scales. Inflation also has the merit of providing a way to preserve primordial inhomogeneities, the power spectrum of which is currently being probed in several experiments~\cite{Ade:2015tva,Ade:2015xua,Ade:2015lrj,Array:2015xqh}. Although the precise mechanism driving the inflationary dynamics is still a mystery, these observations have started to constrain the properties of known inflationary models. In particular, the latest data from the BICEP2/Keck collaboration~\cite{Array:2015xqh} cast strong constraints on the tensor-to-scalar ratio $r$, a quantity related to the amplitude of primordial gravitational waves and to the scale of inflation. As a consequence, the predictions of the so-called ``linear inflation'' model for $r$ as a function of the scalar spectral index $n_s$ now lie about $2 \sigma$ away from the central value, leaving linear inflation as the first model to be possibly ruled out with the upcoming data release.

Although the scalar potential of linear inflation may seem unusual within a quantum field theory context, it has been shown that Coleman-Weinberg (CW) inflation~\cite{Coleman:1973jx} can give rise to a linear potential provided that the inflaton field is non-minimally coupled to gravity and that the Planck scale is dynamically generated~\cite{Rinaldi:2015yoa,Kannike2016,Barrie:2016rnv,Artymowski:2016dlz,Racioppi2017}\footnote{An alternative mechanism relying on fermion condensates is presented in Ref.~\cite{Iso:2014gka}.}. More in general, CW inflation is a well established framework~\cite{Linde:1981mu,Albrecht:1982wi,Linde:1982zj,Ellis:1982ws,Ellis:1982dg} which recently became the subject of a new wave of studies~\cite{Kannike:2014mia,Kannike:2015apa,Marzola:2015xbh,Marzola:2016xgb,Kannike:2016wuy,
Rinaldi:2015uvu,Farzinnia:2015fka,Karananas:2016kyt,Tambalo:2016eqr,Kaneta:2017lnj,Saadi:2017dns} motivated by the discovery of the Higgs boson at the LHC~\cite{Aad:2012tfa,Chatrchyan:2012xdj} and by the apparent lack of symmetries that stabilise its mass. These studies embrace the principle of classical scale invariance as a possible answer~\cite{Bardeen:1995kv,Heikinheimo:2013fta}, proposing a dynamical generation of mass scales via dimensional transmutation \cite{Coleman:1973jx}, through CW potentials induced by new scalar particles~\cite{Hempfling:1996ht,Gabrielli:2013hma,Marzola:2015xbh,Marzola:2016xgb,Marzola:2017jzl}. 

In the present paper, we remain within the framework of CW linear inflation, studying precisely the model where the inflaton is coupled to gravity non-minimally~\cite{Bezrukov:2007ep,Bezrukov:2010jz} and the dynamics is regulated by a CW potential. The novelty of our work lies is in the comparison of the results obtained in the slow-roll formalism with the predictions of constant roll~\cite{Martin2013a, Motohashi2015a, Davydov2016, Motohashi2017a, Gao2017, Odintsov2017, Odintsov2017a, Nojiri2017, Motohashi2017b, Gao2017a, Oikonomou2017, Odintsov2017b, Oikonomou2017a,  Cicciarella2018, Awad2017, Anguelova2017, Ito2017, Yi2017, Mohammadi2018, Gao2018a, Gao2018}, which generalises the former. Whereas in the slow-roll regime the second-derivative term in the Klein-Gordon equation of the inflaton is simply neglected, within constant roll it is taken to be proportional to the first derivative of the field, in a way that the rate of rolling of the inflaton field, $\ddot\phi/(H\dot\phi) \equiv \beta$, is constant. Depending on the value of $\beta$, constant roll then interpolates between the standard slow-roll regime, recovered for $\beta \simeq 0$, and the ultra-slow-roll case~\cite{Anguelova2016, Anguelova2016a, Awad2017, Cai2016, Gong2017, Hirano2016, Martin2013a, Tsamis2004} given by $\beta=-3$. Not all these solutions comply with the current observational bounds, in particular models with $\beta<0$ induce an anomalous super-Hubble evolution of curvature perturbations that place the scale of inflation below the big bang nucleosynthesis one~\cite{Motohashi2015a}. As we will see, for the case of linear inflation, the constant-roll formalism selects a region of the parameters space which respects the current experimental bounds and that results in predictions well distinguished from the corresponding slow-roll solutions of the model.    

The paper is organised as follows: in Sec.~\ref{sec:CR} we briefly revise the basis of constant-roll inflation, before introducing the specific model under examination in Sec.~\ref{sec:model}. In Sec.~\ref{sec:Results} we then illustrate the results obtained for the relevant inflationary observables under both the slow-roll and constant-roll regimes. We conclude in Sec.~\ref{sec:Summary} and present further technical details in Appendix~\ref{sec:appendix}.

\section{Constant-roll inflation} \label{sec:CR}
Constant-roll inflation is a class of phenomenological models characterised by a constant rate of the inflaton~\cite{Martin2013a, Motohashi2015a, Oikonomou2017a, Anguelova2017, Awad2017, Cicciarella2018, Davydov2016, Gao2017a, Gao2017, Ito2017, Mohammadi2018, Motohashi2017a, Motohashi2017b, Nojiri2017, Odintsov2017, Odintsov2017a, Odintsov2017b, Oikonomou2017, Yi2017, Gao2018, Gao2018a}
\begin{equation}
\frac{\ddot{\phi}}{H \dot{\phi}} = \beta \,,
 \label{eq:CR}
\end{equation}
with $\beta$ being a constant. The framework interpolates between the slow-roll inflation, for which $\ddot\phi\simeq0$ and the so-called ultra-slow-roll inflation~\cite{Hirano2016,Cai2016}, satisfying\footnote{Throughout the paper a prime will denote differentiation with respect to the inflaton field $\phi$, while a dot will stand for differentiation with respect to cosmic time $t$. The metric signature we adopt is $\{+, -, -, -\}$.} $ V'(\phi) = 0$ over a range of field values. These two regimes are respectively reproduced for $\beta \simeq 0$ and $\beta = -3$. 

The \textit{Hubble} slow-roll parameters (HSRPs) are defined as 
\begin{align}
\epsh &= 
	2\mpl^2\left(\frac{H'(\phi)}{H(\phi)}\right)^2 
	= - \frac{\dot{H}}{H^2} \,, \\
\etah & 
	= 2\mpl^2 \frac{H''(\phi)}{H(\phi)}  
	= - \frac{ \ddot{\phi}}{ H \,\dot{\phi}} \equiv - \beta \, .
\end{align}
These quantities generally differ from the \textit{potential} slow-roll parameters (PSRPs), which have the following forms:
\begin{eqnarray}
\epsv &=& \frac{\mpl^2}{2} \left( \frac{V'(\phi)}{V(\phi)} \right)^2 \,, \\
\label{eps_V}
\eta_{\rm V} &=& \mpl^2 \, \frac{V''(\phi)}{V(\phi)} \, ,
\label{eta_V}
\end{eqnarray}
with $V$ being the Einstein frame scalar potential.
The Friedmann equations for the problem at hand are
\begin{align}
H^2 & = \frac{1}{3 \mpl^2} \left[ \frac{1}{2} \dot{\phi}^2 + V(\phi) \right] \,,
\label{EOM1A_EF}
\\
\dot{H} & = - \frac{1}{2 \mpl^2} \dot{\phi}^2 \,,
\label{EOM2A_EF}
\end{align}
whereas the inflaton equation of motion is 
\begin{equation}
\ddot{\phi} + 3 H \dot{\phi} + V'(\phi) = 0\, .
\label{EOM3A_EF}
\end{equation}
By using these equations we can now express the first two PSRPs in terms of the corresponding HSRPs as~\cite{Liddle1994}
\begin{align}
\epsv &= \epsh \left( \frac{3 - \etah}{3 - \epsh} \right)^2\,,
\label{system1}
\\
\etav &= \sqrt{2\mpl^2\,\epsh}\, \frac{\eta'_{\rm H}}{3 - \epsh} + \left( \frac{3 - \etah}{3 - \epsh} \right) \left( \epsh + \etah  \right)\,.
\label{system2}
\end{align}
Clearly, in the slow-roll approximation ($\epsh \ll 1$, $\beta\to 0$) we simply have~\cite{Liddle1994}
\begin{align}
\epsv &\simeq \epsh \,, 
\label{epsHSR}
\\
\etav &\simeq \etah+\epsh\,. 
\label{etaHSR}
\end{align}
Differently, for constant roll the first term in the RHS of eq.~\eqref{system2} vanishes identically and it is therefore possible to solve the coupled system of equations to find expressions for $\epsh$ and $\etah$ in terms of the PSRPs; the full solution is presented in Appendix~\ref{sec:appendix}.
Notice that at the end of inflation, when $\epsh = 1$, from eq.~\eqref{system1} it follows that
\begin{equation}
\etah = 3 - 2 \sqrt{\epsv} \,.
\end{equation}

We also remark that in the context of constant roll, $\etah$ remains constant for the whole duration of inflation. Consequently, if we specify the value of the scalar field $\phi$ at the end of inflation by means of eq.~\eqref{epsHCR}, we can easily calculate $\epsv$ and then $\etah$ (or $\beta$). 

To conclude this technical introduction, we report the standard expressions for the tensor-to-scalar ratio $r$ and the scalar spectral index $n_s$ in terms of the HSRPs:
\begin{align}
 r &= 16 \epsh \,,\\
 n_s &= 1 -4\epsh + 2\etah\,. 	
\end{align}

\section{The model} \label{sec:model}

Linear inflation and its different realisations~\cite{Rinaldi:2015yoa, Kannike2016, Barrie:2016rnv, Artymowski:2016dlz, Racioppi2017} recently took the spot because the predictions of the slow-roll formalism within this model fall on the outer boundary of the 2-$\sigma$ region indicated by the BICEP2 data~\cite{Array:2015xqh}. It is then plausible that linear inflation will be ruled out, if not confirmed, with the next data release. In preparation for the latter, it is then important to detail the theoretical aspects of the theory to fully characterise the scenario. In regard of this, and according to our knowledge, we present below the first analysis of the constant-roll predictions within linear inflation. 

In particular, we study the effects of constant-roll inflation within the  non-minimally coupled model presented originally in~\cite{Kannike2016}, where linear inflation appears as an attractor solution. The model is encapsulated in the following Lagrangian, given in the Jordan frame (denoted by barred quantities): 
\begin{equation}
  \sqrt{- g} \mathcal{L} = \sqrt{- g} \left[ - \frac{\xi_\phi}{2} \phib^{2} R 
  + \frac{(\partial \phib)^{2}}{2} - \vb_{\rm 1-loop}(\phib) +  \Lambda^4  \right]\, .
   \label{eq:Jordan:Lagrangian}
\end{equation}
Here $R$ is the Ricci scalar, $\vb_{\rm 1-loop}(\phib)$ a generic 1-loop CW scalar potential,
 \begin{equation}
  \vb_{\rm 1-loop}(\phib) = \frac{1}{4} \left( \lambda_\phi(v_{\phi}) + \beta_{\lambda _{\phi}}(v_{\phi}) \ln\frac{\phib }{v_{\phi}} \right) \phib^4 \label{eq:Veff:J},
 \end{equation}
 and $v_{\phi}$ is the vacuum expectation value (VEV) of the inflaton $\bar\phi$. 
The latter is also responsible for inducing an Einstein-Hilbert term in the Lagrangian~\eqref{eq:Jordan:Lagrangian}. Hence, in order to dynamically generate the Planck scale, the VEV $v_{\phi}$ of the inflaton field must be set to
\begin{equation}
  v_{\phi}^{2} = \frac{\mpl^{2}}{\xi_\phi}.
  \label{eq:v:phi:Planck:mass}
\end{equation}
Notice that such a relation automatically implies $\xi_\phi>0$.

As is customary in inflation model building, the cosmological constant $\Lambda$ is tuned so that the potential vanishes at its minimum:
 \begin{equation}
  \vb_\text{eff}(v_{\phi})= \vb_{\rm 1-loop}(v_{\phi}) + \Lambda^4=\frac{1}{4} \lambda_\phi (v_{\phi}) v_{\phi}^4 + \Lambda^4 = 0 \label{eq:Vmin} \, .
 \end{equation}
By using eqs.~\eqref{eq:v:phi:Planck:mass} and \eqref{eq:Vmin}, as well as by requiring that $v_{\bar\phi} \neq0$, it is possible to show~\cite{Kannike2016} that the inflaton potential can be rewritten as
\begin{eqnarray}
  \vb_{\rm eff}(\phi)&=& \frac{1}{4} \lambda_\phi (\phib) \phib^4 + \Lambda^4 =  \Lambda ^4 \left\{ 1 + \left[ 2 \ln \left(\frac{\xi_\phi \phib^2}{\mpl^2}\right) -1 \right] \frac{\xi_\phi^2 \phib^4}{\mpl^4} \right\} \label{eq:Veff:Jordan:Lambda}\,.
\end{eqnarray}

Choosing now to work in the Einstein frame for simplicity\footnote{An analysis of the invariant approach along the lines of~\cite{Jaerv2015, Kuusk2016, Kuusk2016a, Jaerv2017, Karam2017} is postponed to a future work.},
the scalar potential is transformed to 
\begin{equation}
  V(\phi) = \frac{\mpl^4}{\xi_\phi^2 \phib^4} \vb_{\rm eff}(\phib)
=\Lambda^4 \left(4\sqrt{\frac{\xi_\phi}{1+6 \, \xi_\phi}} \frac{\phi}{\mpl}+e^{-4\sqrt{\frac{\xi_\phi}{1+6 \, \xi_\phi}} \frac{\phi}{\mpl}}-1\right) \, , \label{eq:Veff:Einstein}
\end{equation}
where the canonically-normalised field $\phi$ is related to the Jordan frame field $\phib$ via
\begin{equation}
\phi = \sqrt{\frac{1+6 \, \xi_\phi}{\xi_\phi }} \mpl \ln \frac{\phib}{v_{\bar\phi}} \, .
\label{eq:phiE}
\end{equation}
The Einstein frame potential then depends formally only on two parameters: $\Lambda$, and $\xi_\phi$. As $\Lambda$ will be employed to satisfy the constraint on the amplitude of scalar perturbations \cite{Ade:2015lrj,Ade:2015xua}, the values obtained for the (normalisation-independent) observables $r$ and $n_s$ will depend only on the value of $\xi_\phi$.

\subsection{Reaching the constant roll limit} 
\label{sub:Realising the constant roll dynamics}

Before presenting the results obtained with the model at hand, we show that the constant roll dynamics is indeed achieved and maintained by the inflaton during its evolution. 

To this purpose, we compare the field trajectories of the complete Klein-Gordon equation for the inflaton field~\eqref{EOM3A_EF} 
to the ones obtained under the constant-roll approximation,
\begin{equation}
\left( 3 + \beta \right) H \dot{\phi} + V'(\phi) = 0 \, 
\label{Klein-Gordon-full-CR}
\end{equation}
for the potential specified in eq.~\eqref{eq:Veff:Einstein}.

\begin{figure}
  \begin{center}
\includegraphics[width=0.8\linewidth]{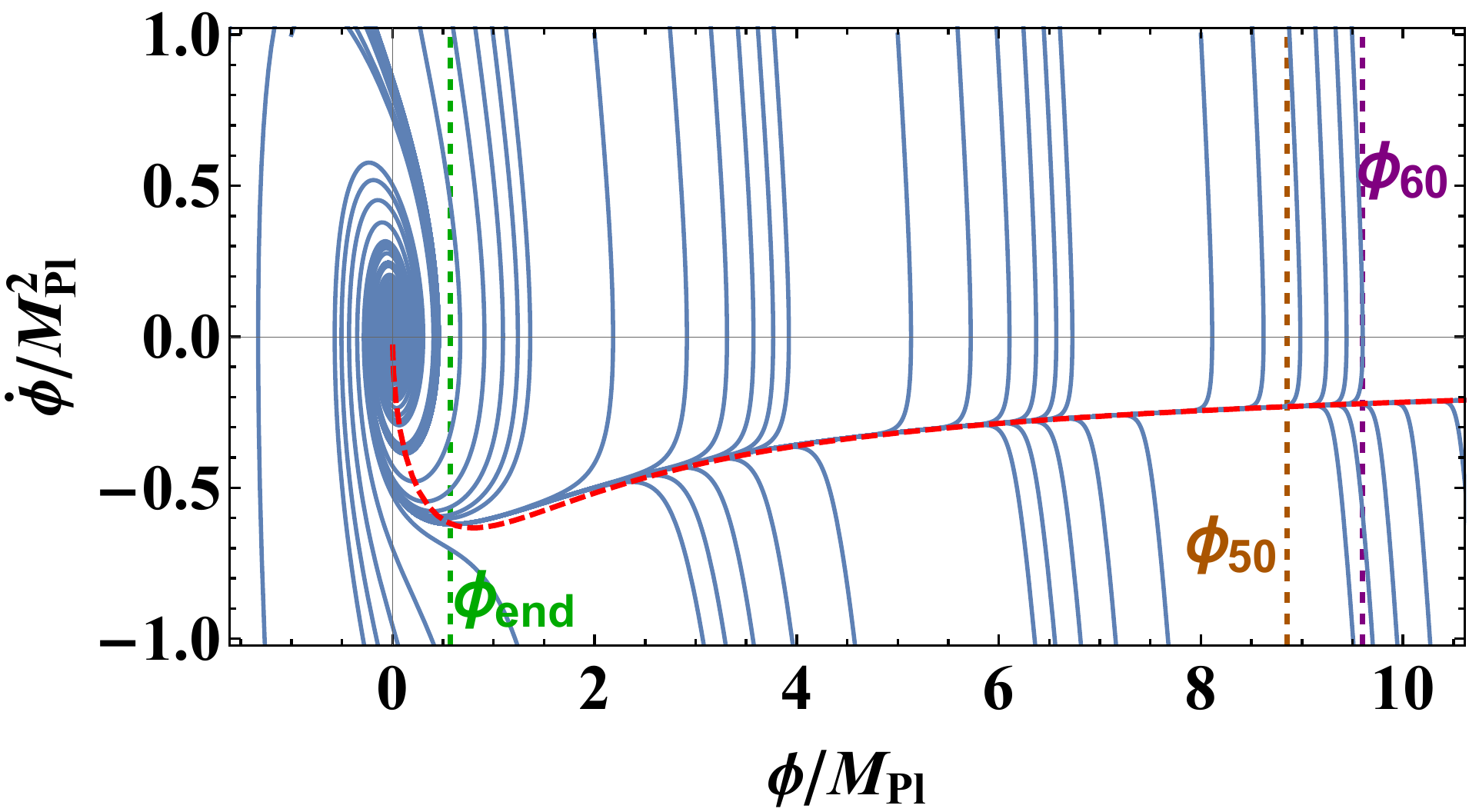}
    \caption{Dynamics of the linear inflation model for $\xi = 0.3$, obtained by numerically solving the Klein-Gordon equation~\eqref{EOM3A_EF} for various initial conditions. The red dashed curve corresponds to the solution of the constant-roll approximation to the Klein-Gordon equation~\eqref{Klein-Gordon-full-CR} with $\beta = 0.005$. The green line shows the value of the scalar field at the end of inflation, while the orange and purple lines indicate the values of the scalar field which reproduce $50$ and $60$ $e$-folds of inflation, respectively.}
   \label{Attractor}
  \end{center}
\end{figure}

In Fig.~\ref{Attractor} we show the results obtained in the $\phi - \dot{\phi}$ phase space for different initial conditions and for $\xi = 0.3$. This value for the non-minimal coupling corresponds to the linear inflation limit~\cite{Kannike2016} of the model under consideration. The red dashed line corresponds to the solution of~\eqref{Klein-Gordon-full-CR} with $\beta = 0.005$ (see also Fig.~\ref{etaHCR}). The attractor behaviour of the constant roll solution is manifest. The exact attractor trajectory is very well approximated by the
constant roll trajectory, especially for $\phi > \mpl$, which is the region that is most relevant in the computation of the inflationary parameters. We find a similar attractor behaviour for other values of $\xi$ and $\beta$ as well. 

The stability of the constant-roll attractor can be further investigated by employing the technique developed in~\cite{Liddle1994}. We can write the scalar potential in terms of $H(\phi)$ as
\begin{equation}
V(\phi) = 3 \mpl^2 H(\phi)^2 - 2 \mpl^4 \left( H'(\phi) \right)^2 \,.
\end{equation}
Then, varying the above equation we have
\begin{equation}
H'_0(\phi) \delta H'(\phi) \simeq \frac{3}{2 \mpl^2} H_0(\phi) \delta H(\phi) \,,
\end{equation}
which, for a given solution $H_0 (\phi)$, becomes
\begin{equation}
\delta H(\phi) = \delta H(\phi_0) e^{\frac{3}{2 \mpl^2} \int_{\phi_0}^\phi \frac{H_0(\phi)}{H'_0(\phi)} \dd \phi} \,,
\end{equation}
with $\phi_0$ some initial value of the canonically-normalised scalar field.

\begin{figure}[H]
  \begin{center}
\includegraphics[width=0.7\linewidth]{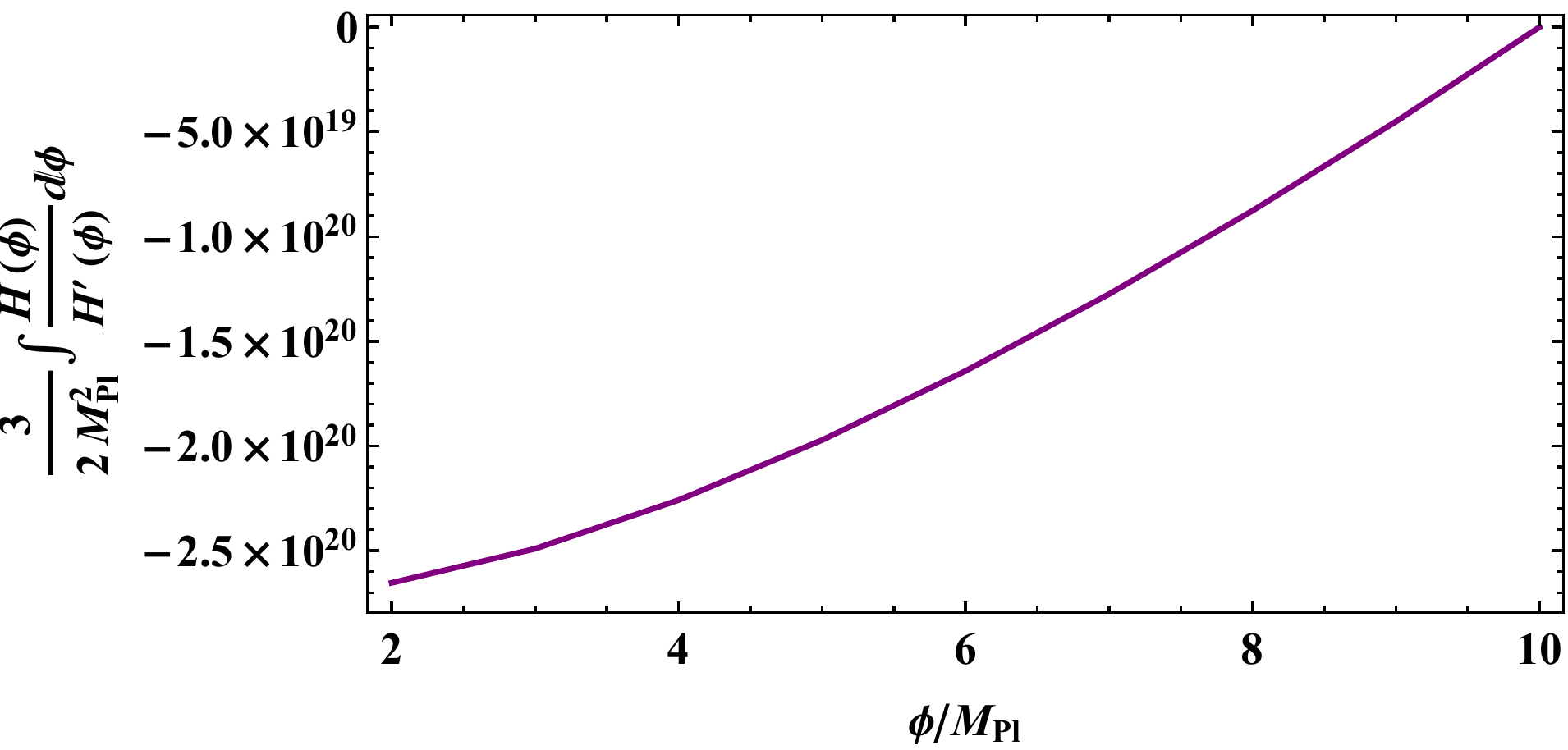}
  \caption{Stability of linear perturbations for the Hubble parameter.}
   \label{Stability}
  \end{center}
\end{figure}

In Fig.~\ref{Stability} we show that the linear perturbations of the Hubble parameter are decaying, which implies that the attractor solutions are stable.

Finally, we note that although the potential \eqref{eq:Veff:Einstein} is not among the exact constant roll solutions found in~\cite{Motohashi2015a}, it can still be considered as an approximately constant roll one in the sense that $\vert \dot{\beta} \vert / (H \vert \beta \vert ) \ll 1$ if $\beta \ll 1$ {\it i.e.} in a slow-roll regime. One can easily verify that this is indeed the case by performing a numerical calculation of
\begin{equation}
\frac{\dot{\beta}}{H \beta}= \frac{3\epsilon_H}{\beta}+\epsilon_H-(3+\beta) -\frac{1}{H^2 \beta} V''(\phi),
\end{equation}
which is obtained after differentiating the Klein-Gordon equation \eqref{Klein-Gordon-full-CR} with respect to $t$ and keeping in mind that in the most general case the parameter $\beta$ may exhibit a non trivial dependence on the inflaton field value. This is also confirmed by our results in the Section \ref{sec:Results} (in particular Figs. \ref{Fig:r_vs_n} and \ref{etaHCR}).

\section{Results} \label{sec:Results}

\begin{figure}[h!]
  \begin{center}
\includegraphics[width=0.65\linewidth]{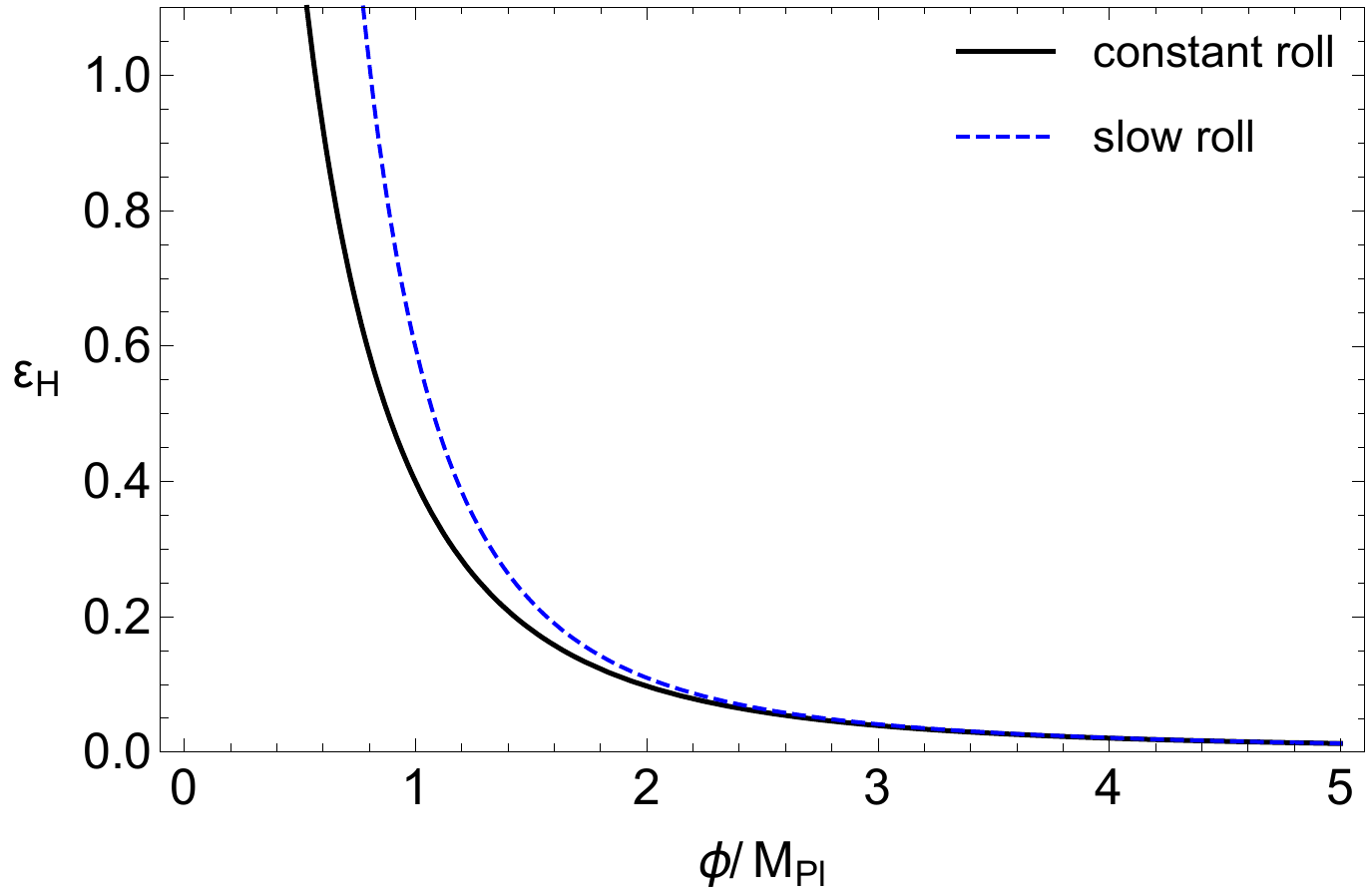}
    \caption{The Hubble parameter $\epsh$ as a function of the Einstein frame scalar field $\phi$ in the constant-roll approximation (black solid line, c.f. eq.~\eqref{epsHCR}) and in the slow roll approximation (blue dashed line, c.f. eq.~\eqref{epsHSR}) for $\xi_\phi=0.3$ and $N=60$ $e$-folds. The difference between the two approximations becomes sizeable near the end of inflation, for $\epsh=1$, and results in different predictions for the observables.}
   \label{EHSRvsEHCR}
  \end{center}
\end{figure}

In Fig.~\ref{EHSRvsEHCR} we plot the first Hubble parameter $\epsh$ as a function of the inflaton $\phi$ in the constant-roll and slow-roll approximations for $N=60$ $e$-folds and $\xi_\phi = 0.3$. The black solid line corresponds to the constant-roll regime, c.f. eq.~\eqref{epsHCR}, while the blue dashed line corresponds to the slow-roll approximation encapsulated in the standard relation of eq.~\eqref{epsHSR}. As we can see, the difference between the two approaches diminishes with larger values of $\phi$ and becomes relevant near the end of inflation, when we have exactly $\epsh=1$. 

\begin{figure}[h!]
  \begin{center}
\includegraphics[width=0.65\linewidth]{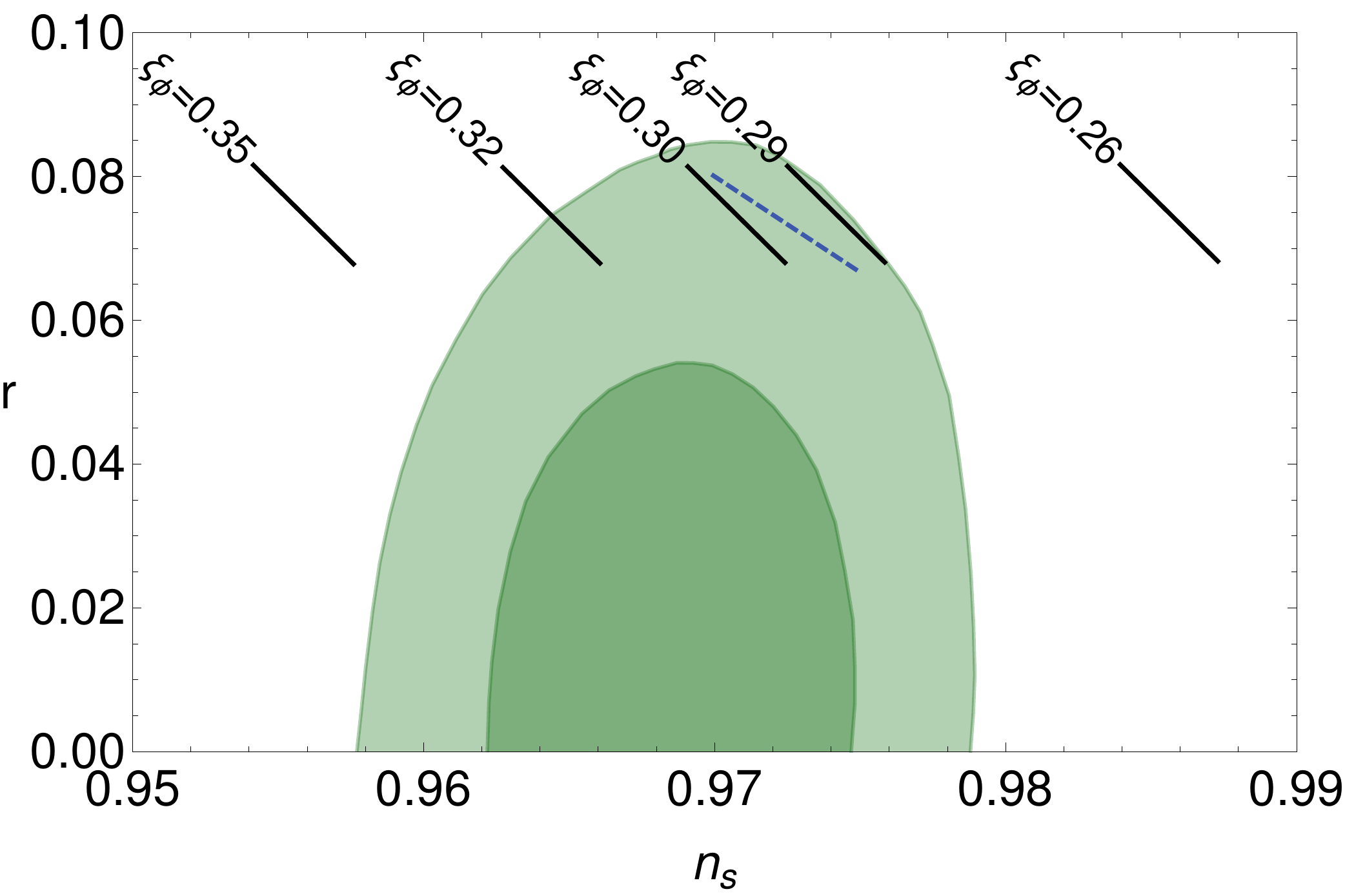}
    \caption{Tensor-to-scalar ratio $r$ as a function of $n_s$ for the 1st order (blue dashed line, $\xi_\phi=0.3$) slow-roll approximation and in constant roll (black solid line) for $N_e \in [50,60]$ $e$-folds. The green areas present the 1 and 2$\sigma$ best fit of the BICEP2/Keck data~\cite{Array:2015xqh}.}
   \label{Fig:r_vs_n}
  \end{center}
\end{figure}

As the discrepancy between slow roll and constant roll is sizeable only in the latest stage of inflation, we expect the latter to deliver similar results for the scalar-to-tensor ratio $r$. This is confirmed in Fig.~\ref{Fig:r_vs_n}, where we plot this quantity as a function of $n_s$ for the 1st order slow-roll approximation (blue dashed line; $\xi_\phi=0.3$)\footnote{In the slow-roll regime the linear limit is already saturated by $\xi_\phi\gtrsim0.3$ \cite{Kannike2016}, therefore we only plot the case $\xi_\phi=0.3$.} and for the constant-roll regime (black solid line, $\xi_\phi = 0.26$, $0.29$, $0.30$, $0.32$, $0.35$), taking $N_e \in [50,60]$ $e$-folds. The (light) green area shows the (1) 2$\sigma$ best fit of the BICEP2/Keck data~\cite{Array:2015xqh}. As we can see, although the predictions for $r$ are remarkably similar ($\Delta r = 0.001$), the two approaches yield substantially different values of the spectral index $n_s$ computed as a function of $\xi_\phi$. In particular, we notice that the 2-$\sigma$ bound from the BICEP data forces $\xi_\phi\sim 0.29-0.32$. 

\begin{figure}[h!]
  \begin{center}
\includegraphics[width=0.65\linewidth]{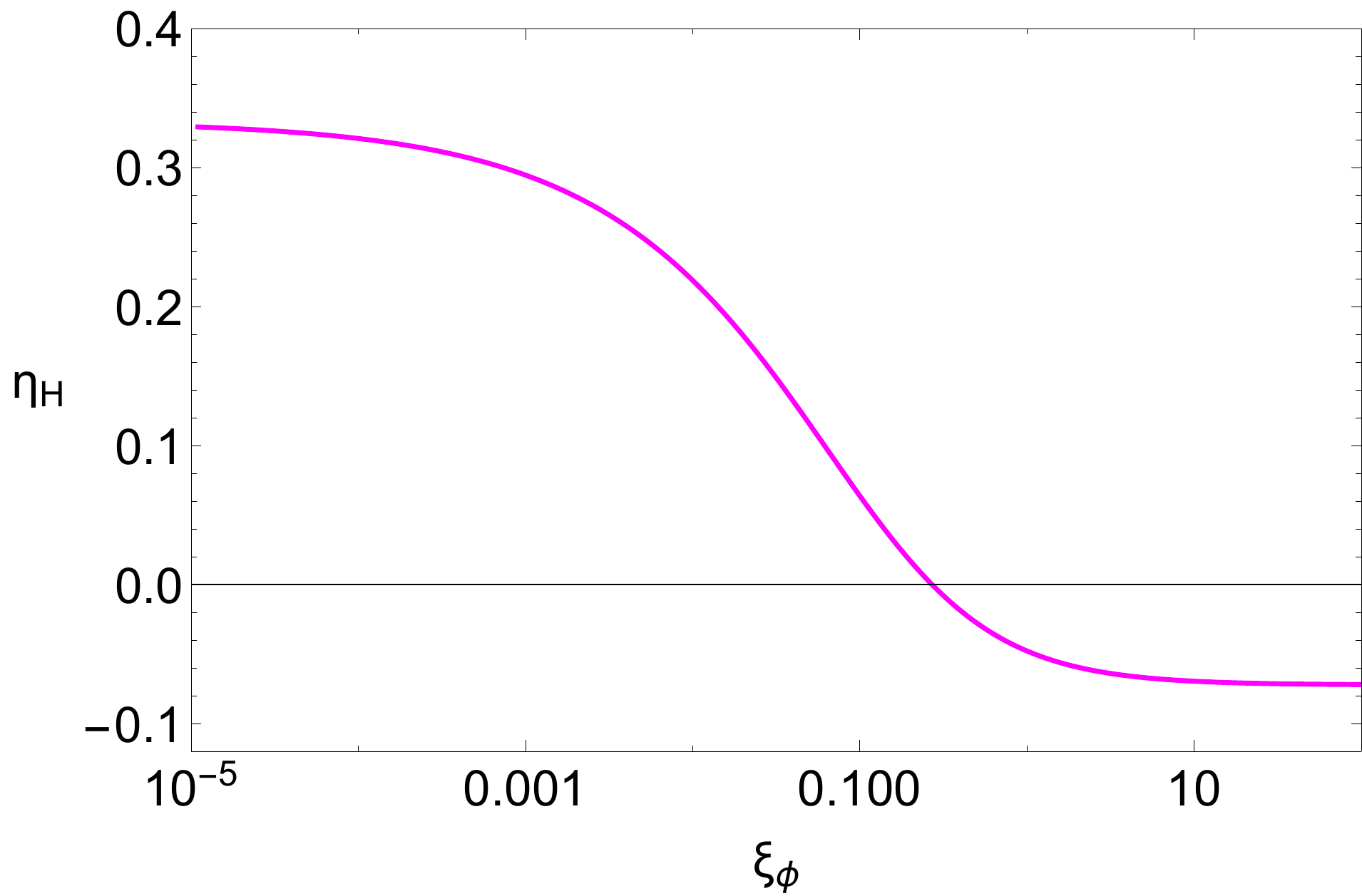}
    \caption{The parameter $\etah$ in the constant-roll approach for $\xi_\phi$ in the range $(10^{-5},10^{2})$.}
   \label{etaHCR}
  \end{center}
\end{figure}

To further investigate the issue, we plot in Fig.~\ref{etaHCR} the constant-roll parameter $\etah \equiv - \beta$ obtained for various values of the non-minimal coupling $\xi_\phi$. For $\xi_\phi\notin[0.29, 0.32]$, we find that $\etah$ assumes values large in magnitude which push the predictions of constant-roll linear inflation into the disfavoured region.

Fig.~\ref{etaHCR} makes also clear that the requirement $\beta\gtrsim 0$, which prevents the evolution of super-Hubble curvature perturbations~\cite{Motohashi2015a}, casts a mild lower bound on the value of the non-minimal coupling $\xi_\phi$ within the present framework. 
Notice also that the curvature perturbations in the superhorizon regime have the form~\cite{Martin2013a} 
\begin{equation}
\zeta_k \propto A_k + B_k a^{- \left( 2 \beta + 3 \right)} \, ,
\label{superhorizon}
\end{equation}
where $a$ is the scale factor. As in our case the constant-roll parameter $\beta = -\eta_H\in\left[ - 0.325 , 0.07 \right]$, see Fig.~\ref{etaHCR}, the second term in the above equation is suppressed and the scenario has no superhorizon evolution of the curvature perturbations. 

\begin{figure}[h!]
  \begin{center}
\includegraphics[width=0.65\linewidth]{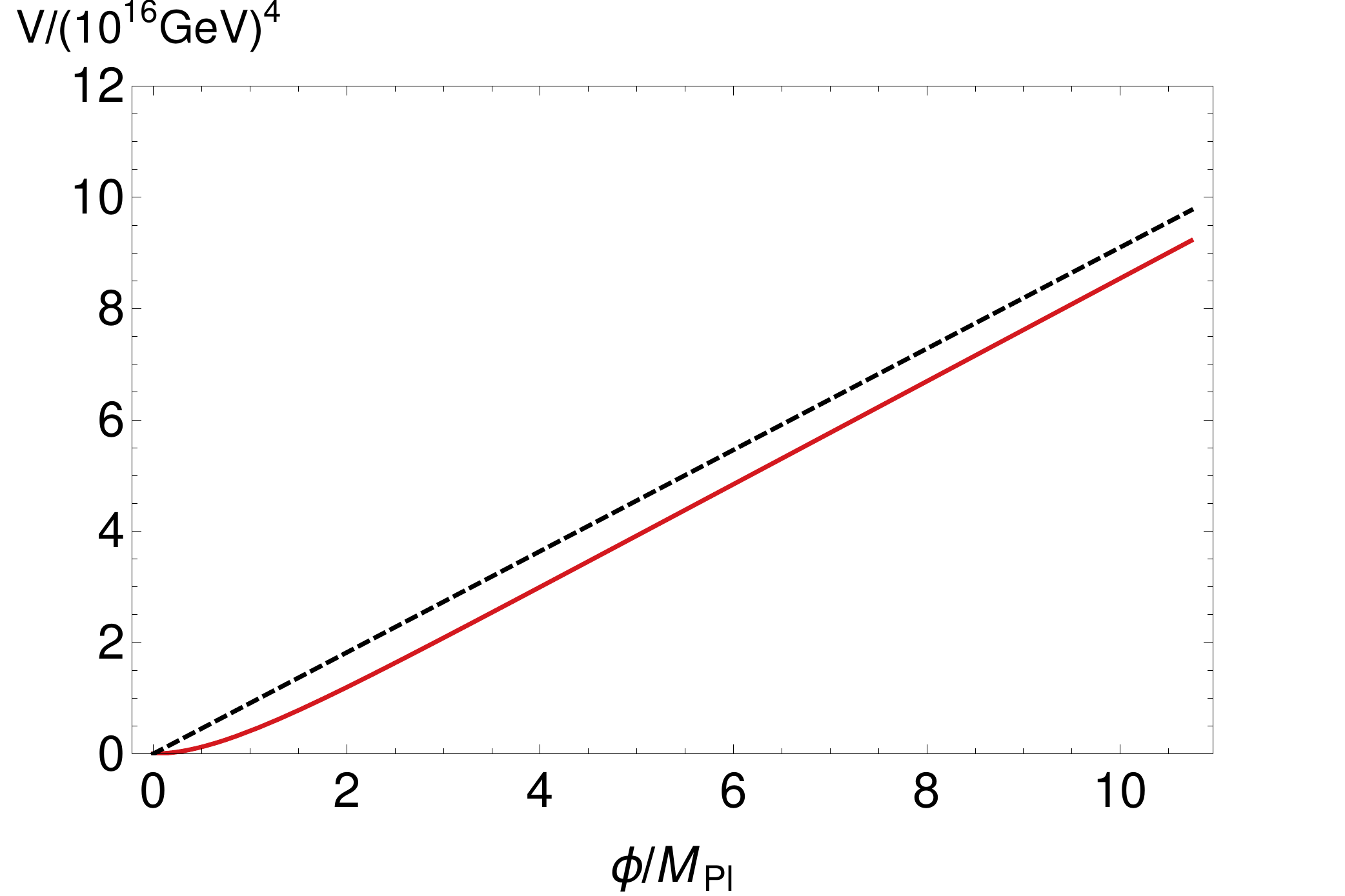}
    \caption{Inflaton potential $V$ (red line) as a function of the field value $\phi$ for fixed $\Lambda \simeq 9.2 \times 10^{15}$ GeV and $\xi_\phi=0.3$. The black dashed line shows the linear limit.}
   \label{Fig:V_shape}
  \end{center}
\end{figure}

To conclude, we remark that the exact constant-roll linear inflation limit, corresponding to  $\xi_\phi \to \infty$, is ruled out by the BICEP2/Keck data. Because only quasi-linear potentials like the one depicted in Fig.~\ref{Fig:V_shape} are allowed, we named the proposed scenario \emph{constant-roll quasi-linear inflation}.

\section{Conclusions} \label{sec:Summary}

The slow-roll approximation is generally considered to be an adequate description of the dynamics of inflation. The expected precision of future experiments, however, prompts us to examine the equation of motion of the inflaton in more detail. To this end, in the present paper we considered constant-roll inflation, a scenario in which the inflaton rolls down its potential at a constant rate throughout the whole inflationary process. By expressing the inflaton equation of motion in terms of the first two Hubble ($\epsh$, $\etah$) and potential ($\epsilon_V$, $\eta_V$) slow-roll parameters, we found analytical solutions of the former in terms of the latter. With such expressions, for a given potential we can compute the value of the inflaton field at the end of inflation by setting $\epsh=1$, similarly to the usual slow-roll case. The values for the relevant inflationary observables can then be obtained by requiring a duration of $N=50-60$ $e$-folds. 

Since linear inflation is presently at the boundary of the region allowed by the experiments, we chose to apply our method to the non-minimal Coleman-Weinberg model developed in~\cite{Kannike2016}, presenting an Einstein frame potential which interpolates between quadratic and linear inflation. We found that at a given number of inflation $e$-folds the tensor-to-scalar ratio differs by about $\Delta r = 0.001$ in the two approaches. Such a discrepancy is within the sensitivity of the CORE mission \cite{Remazeilles:2017szm}. Furthermore, we found that in the constant-roll regime $\etah$, and consequently the scalar index $n_s$, strongly depends on the value of the non-minimal coupling $\xi_\phi$. The current 2-$\sigma$ confidence interval of $n_s$ then constrains the non-minimal coupling to be $\xi_\phi \sim 0.29-0.31$, preventing the model from reaching the linear limit valid for $\xi_\phi \to \infty$.  

\acknowledgments

L.M. and A.R. are supported by the Estonian Research Council grants IUT23-6, PUT1026 and by the ERDF Centre of Excellence project TK133. T.P. would like to thank the Alexander S. Onassis Public Benefit Foundation for financial support. We also thank the referee for their insightful comments and suggestions.

\appendix

\section{Connection between potential and Hubble slow-roll parameters in the constant-roll approximation} 
\label{sec:appendix}
We present here the full solutions of the algebraic system of equations in eqs.~\eqref{system1} and~\eqref{system2} obtained for a constant value of $\eta_H$. 

For $\epsh$ we find
\begin{equation}
\epsh =\frac{1}{3 \epsv}\left[ B^2-9 \epsv^2 \left(3+2 \epsv-2\etav \right) \sqrt[3]{\frac{2}{A}}+  \sqrt[3]{\frac{A}{2}} B \right] \,,
\label{epsHCR}
\end{equation}
where
\begin{equation}
\begin{split}
A \equiv \, & 27 \epsv^2 \sqrt{-4 \epsv^3+12 \epsv^2 (\etav+11)-12 \epsv \left(\etav^2+4 \etav-3\right)+\etav^2 (4 \etav-3)} \left| \etav-2 \epsv \right| \\
&+2 \epsv^6-6 \epsv^5 (2 \etav+15)+6 \epsv^4 \left(5 \etav^2+51 \etav+117\right)-2 \epsv^3 \left(20 \etav^3+189 \etav^2+297 \etav-27\right) \\ 
&+3 \epsv^2 \etav^2 \left(10 \etav^2+66 \etav+45\right)-12 \epsv \etav^4 (\etav+3)+2 \etav^6
\end{split}
\end{equation}
and
\begin{equation}
B \equiv \left( \epsv^2-2 \epsv \left( \etav +3 \right)+\etav^2 \right)\,.
\end{equation}
As for $\etah$, we have 
\begin{equation}
\begin{split}
\eta_H &= \frac{6-\eps_V+\eta_V}{3}+ C \frac{2^{2/3} \eta_V^3- (2 D)^{1/3} \eta_V-2^{5/3} \eps_V \eta_V (3+\eta_V)+2^{2/3} \eps_V^2 (18+\eta_V)}{2 D^{2/3} \eps_V^2 (2 \eps_V-\eta_V)}
    \\
&\quad 
+\frac{2^{1/3} \left[ 6 \eps_V^2 (10+\eta_V)-2 \eps_V^3+\eta_V^2 (3+2 \eta_V)-6 \eps_V (3+6 \eta_V+ \eta_V^2) \right]}{6 D^{1/3}}
    \\
&\quad 
+\frac{1}{3}\frac{10 \eps_V^4 (9+\eta_V)-2\eps_V^5+\eta_V^4 (3+2\eta_V)-2 \eps_V \eta_V^2 \left( 18+27 \eta_V +5 \eta_V^2  \right)}{2^{1/3} D^{2/3}}
    \\
&\quad 
+\frac{1}{3}\frac{\eps_V^2 \left[414\eta_V+189 \eta_V^2+20 \eta_V^3-2\eps_V \left(351+114 \eta_V+10 \eta_V^2 \right)-54\right]}{2^{1/3} D^{2/3}} \,,
\end{split}
\end{equation}

where we have defined the following quantities:

\begin{equation}
C \equiv \eps_V^2 \sqrt{(-2 \eps_V+\eta_V)^2 \left[ 12 \eps_V^2 (11+\eta_V)-4 \eps_V^3+\eta_V^2 \left(4 \eta_V-3 \right)-12 \eps_V \left( 4 \eta_V+\eta_V^2-3 \right) \right]}
\end{equation}
and
\begin{equation}\
\begin{split}
D & \equiv 27 \, C +2 \eps_V^6+2 \eta_V^6-12 \eps_V  \eta_V^4 (3+\eta_V)-6 \eps_V^5 (15+2 \eta_V)
    \\
&\quad 
+6 \eps_V^4 (117+51 \eta_V+5 \eta_V^2)+3 \eps_V^2 \eta_V^2 (45+66 \eta_V+10 \eta_V^2)
    \\
&\quad 
-2 \eps_V^3 \left(-27 +297 \eta_V+189 \eta_V^2+20 \eta_V^3 \right) \,.
\end{split}
\end{equation}

\bibliographystyle{JHEP}
\bibliography{References}

\providecommand{\href}[2]{#2}\begingroup\raggedright\begin{thebibliography}{10}

\bibitem{Starobinsky:1980te}
A.~A. Starobinsky, {\it {A New Type of Isotropic Cosmological Models Without
  Singularity}},  {\em Phys. Lett.} {\bf B91} (1980) 99--102.

\bibitem{Guth:1980zm}
A.~H. Guth, {\it {The Inflationary Universe: A Possible Solution to the Horizon
  and Flatness Problems}},  {\em Phys. Rev.} {\bf D23} (1981) 347--356.

\bibitem{Linde:1981mu}
A.~D. Linde, {\it {A New Inflationary Universe Scenario: A Possible Solution of
  the Horizon, Flatness, Homogeneity, Isotropy and Primordial Monopole
  Problems}},  {\em Phys.Lett.} {\bf B108} (1982) 389--393.

\bibitem{Albrecht:1982wi}
A.~Albrecht and P.~J. Steinhardt, {\it {Cosmology for Grand Unified Theories
  with Radiatively Induced Symmetry Breaking}},  {\em Phys. Rev. Lett.} {\bf
  48} (1982) 1220--1223.

\bibitem{Ade:2015tva}
{\bf BICEP2, Planck} Collaboration, P.~A.~R. Ade et~al., {\it {Joint Analysis
  of BICEP2/$Keck Array$ and $Planck$ Data}},  {\em Phys. Rev. Lett.} {\bf 114}
  (2015) 101301, [\href{http://arxiv.org/abs/1502.00612}{{\tt
  arXiv:1502.00612}}].

\bibitem{Ade:2015xua}
{\bf Planck} Collaboration, P.~A.~R. Ade et~al., {\it {Planck 2015 results.
  XIII. Cosmological parameters}},  {\em Astron. Astrophys.} {\bf 594} (2016)
  A13, [\href{http://arxiv.org/abs/1502.01589}{{\tt arXiv:1502.01589}}].

\bibitem{Ade:2015lrj}
{\bf Planck} Collaboration, P.~A.~R. Ade et~al., {\it {Planck 2015 results. XX.
  Constraints on inflation}},  {\em Astron. Astrophys.} {\bf 594} (2016) A20,
  [\href{http://arxiv.org/abs/1502.02114}{{\tt arXiv:1502.02114}}].

\bibitem{Array:2015xqh}
{\bf BICEP2, Keck Array} Collaboration, P.~A.~R. Ade et~al., {\it {Improved
  Constraints on Cosmology and Foregrounds from BICEP2 and Keck Array Cosmic
  Microwave Background Data with Inclusion of 95 GHz Band}},  {\em Phys. Rev.
  Lett.} {\bf 116} (2016) 031302, [\href{http://arxiv.org/abs/1510.09217}{{\tt
  arXiv:1510.09217}}].

\bibitem{Coleman:1973jx}
S.~R. Coleman and E.~J. Weinberg, {\it {Radiative Corrections as the Origin of
  Spontaneous Symmetry Breaking}},  {\em Phys. Rev.} {\bf D7} (1973)
  1888--1910.

\bibitem{Rinaldi:2015yoa}
M.~Rinaldi, L.~Vanzo, S.~Zerbini, and G.~Venturi, {\it {Inflationary
  quasiscale-invariant attractors}},  {\em Phys. Rev.} {\bf D93} (2016) 024040,
  [\href{http://arxiv.org/abs/1505.03386}{{\tt arXiv:1505.03386}}].

\bibitem{Kannike2016}
K.~Kannike, A.~Racioppi, and M.~Raidal, {\it {Linear inflation from quartic
  potential}},  {\em JHEP} {\bf 01} (2016) 035,
  [\href{http://arxiv.org/abs/1509.05423}{{\tt arXiv:1509.05423}}].

\bibitem{Barrie:2016rnv}
N.~D. Barrie, A.~Kobakhidze, and S.~Liang, {\it {Natural Inflation with Hidden
  Scale Invariance}},  {\em Phys. Lett.} {\bf B756} (2016) 390--393,
  [\href{http://arxiv.org/abs/1602.04901}{{\tt arXiv:1602.04901}}].

\bibitem{Artymowski:2016dlz}
M.~Artymowski and A.~Racioppi, {\it {Scalar-tensor linear inflation}},  {\em
  JCAP} {\bf 1704} (2017), no.~04 007,
  [\href{http://arxiv.org/abs/1610.09120}{{\tt arXiv:1610.09120}}].

\bibitem{Racioppi2017}
A.~Racioppi, {\it {Coleman-Weinberg linear inflation: metric vs. Palatini
  formulation}},  {\em JCAP} {\bf 1712} (2017), no.~12 041,
  [\href{http://arxiv.org/abs/1710.04853}{{\tt arXiv:1710.04853}}].

\bibitem{Iso:2014gka}
S.~Iso, K.~Kohri, and K.~Shimada, {\it {Small field Coleman-Weinberg inflation
  driven by a fermion condensate}},  {\em Phys. Rev.} {\bf D91} (2015), no.~4
  044006, [\href{http://arxiv.org/abs/1408.2339}{{\tt arXiv:1408.2339}}].

\bibitem{Linde:1982zj}
A.~D. Linde, {\it {Coleman-Weinberg Theory and a New Inflationary Universe
  Scenario}},  {\em Phys.Lett.} {\bf B114} (1982) 431.

\bibitem{Ellis:1982ws}
J.~R. Ellis, D.~V. Nanopoulos, K.~A. Olive, and K.~Tamvakis, {\it {PRIMORDIAL
  SUPERSYMMETRIC INFLATION}},  {\em Nucl.Phys.} {\bf B221} (1983) 524.

\bibitem{Ellis:1982dg}
J.~R. Ellis, D.~V. Nanopoulos, K.~A. Olive, and K.~Tamvakis, {\it {Fluctuations
  in a Supersymmetric Inflationary Universe}},  {\em Phys.Lett.} {\bf B120}
  (1983) 331.

\bibitem{Kannike:2014mia}
K.~Kannike, A.~Racioppi, and M.~Raidal, {\it {Embedding inflation into the
  Standard Model - more evidence for classical scale invariance}},  {\em JHEP}
  {\bf 06} (2014) 154, [\href{http://arxiv.org/abs/1405.3987}{{\tt
  arXiv:1405.3987}}].

\bibitem{Kannike:2015apa}
K.~Kannike, G.~H{\"u}tsi, L.~Pizza, A.~Racioppi, M.~Raidal, A.~Salvio, and
  A.~Strumia, {\it {Dynamically Induced Planck Scale and Inflation}},  {\em
  JHEP} {\bf 05} (2015) 065, [\href{http://arxiv.org/abs/1502.01334}{{\tt
  arXiv:1502.01334}}].

\bibitem{Marzola:2015xbh}
L.~Marzola, A.~Racioppi, M.~Raidal, F.~R. Urban, and H.~Veerm{\"a}e, {\it
  {Non-minimal CW inflation, electroweak symmetry breaking and the 750 GeV
  anomaly}},  {\em JHEP} {\bf 03} (2016) 190,
  [\href{http://arxiv.org/abs/1512.09136}{{\tt arXiv:1512.09136}}].

\bibitem{Marzola:2016xgb}
L.~Marzola and A.~Racioppi, {\it {Minimal but non-minimal inflation and
  electroweak symmetry breaking}},  {\em JCAP} {\bf 1610} (2016), no.~10 010,
  [\href{http://arxiv.org/abs/1606.06887}{{\tt arXiv:1606.06887}}].

\bibitem{Kannike:2016wuy}
K.~Kannike, M.~Raidal, C.~Spethmann, and H.~Veerm{\"a}e, {\it {The evolving
  Planck mass in classically scale-invariant theories}},  {\em JHEP} {\bf 04}
  (2017) 026, [\href{http://arxiv.org/abs/1610.06571}{{\tt arXiv:1610.06571}}].

\bibitem{Rinaldi:2015uvu}
M.~Rinaldi and L.~Vanzo, {\it {Inflation and reheating in theories with
  spontaneous scale invariance symmetry breaking}},  {\em Phys. Rev.} {\bf D94}
  (2016), no.~2 024009, [\href{http://arxiv.org/abs/1512.07186}{{\tt
  arXiv:1512.07186}}].

\bibitem{Farzinnia:2015fka}
A.~Farzinnia and S.~Kouwn, {\it {Classically scale invariant inflation,
  supermassive WIMPs, and adimensional gravity}},  {\em Phys. Rev.} {\bf D93}
  (2016), no.~6 063528, [\href{http://arxiv.org/abs/1512.05890}{{\tt
  arXiv:1512.05890}}].

\bibitem{Karananas:2016kyt}
G.~K. Karananas and J.~Rubio, {\it {On the geometrical interpretation of
  scale-invariant models of inflation}},  {\em Phys. Lett.} {\bf B761} (2016)
  223--228, [\href{http://arxiv.org/abs/1606.08848}{{\tt arXiv:1606.08848}}].

\bibitem{Tambalo:2016eqr}
G.~Tambalo and M.~Rinaldi, {\it {Inflation and reheating in scale-invariant
  scalar-tensor gravity}},  {\em Gen. Rel. Grav.} {\bf 49} (2017), no.~4 52,
  [\href{http://arxiv.org/abs/1610.06478}{{\tt arXiv:1610.06478}}].

\bibitem{Kaneta:2017lnj}
K.~Kaneta, O.~Seto, and R.~Takahashi, {\it {Very low scale Coleman-Weinberg
  inflation with non-minimal coupling}},
  \href{http://arxiv.org/abs/1708.06455}{{\tt arXiv:1708.06455}}.

\bibitem{Saadi:2017dns}
H.~Saadi, {\it {A Light Dilatonic Higgs and a Dilatonic Inflaton in the
  Georgi-Glashow SU(5) Model}},  \href{http://arxiv.org/abs/1709.01419}{{\tt
  arXiv:1709.01419}}.

\bibitem{Aad:2012tfa}
{\bf ATLAS} Collaboration, G.~Aad et~al., {\it {Observation of a new particle
  in the search for the Standard Model Higgs boson with the ATLAS detector at
  the LHC}},  {\em Phys. Lett.} {\bf B716} (2012) 1--29,
  [\href{http://arxiv.org/abs/1207.7214}{{\tt arXiv:1207.7214}}].

\bibitem{Chatrchyan:2012xdj}
{\bf CMS} Collaboration, S.~Chatrchyan et~al., {\it {Observation of a new boson
  at a mass of 125 GeV with the CMS experiment at the LHC}},  {\em Phys. Lett.}
  {\bf B716} (2012) 30--61, [\href{http://arxiv.org/abs/1207.7235}{{\tt
  arXiv:1207.7235}}].

\bibitem{Bardeen:1995kv}
W.~A. Bardeen, {\it {On naturalness in the standard model}},  in {\em {Ontake
  Summer Institute on Particle Physics Ontake Mountain, Japan, August
  27-September 2, 1995}}, 1995.

\bibitem{Heikinheimo:2013fta}
M.~Heikinheimo, A.~Racioppi, M.~Raidal, C.~Spethmann, and K.~Tuominen, {\it
  {Physical Naturalness and Dynamical Breaking of Classical Scale Invariance}},
   {\em Mod. Phys. Lett.} {\bf A29} (2014) 1450077,
  [\href{http://arxiv.org/abs/1304.7006}{{\tt arXiv:1304.7006}}].

\bibitem{Hempfling:1996ht}
R.~Hempfling, {\it {The Next-to-minimal Coleman-Weinberg model}},  {\em Phys.
  Lett.} {\bf B379} (1996) 153--158,
  [\href{http://arxiv.org/abs/hep-ph/9604278}{{\tt hep-ph/9604278}}].

\bibitem{Gabrielli:2013hma}
E.~Gabrielli, M.~Heikinheimo, K.~Kannike, A.~Racioppi, M.~Raidal, and
  C.~Spethmann, {\it {Towards Completing the Standard Model: Vacuum Stability,
  EWSB and Dark Matter}},  {\em Phys. Rev.} {\bf D89} (2014), no.~1 015017,
  [\href{http://arxiv.org/abs/1309.6632}{{\tt arXiv:1309.6632}}].

\bibitem{Marzola:2017jzl}
L.~Marzola, A.~Racioppi, and V.~Vaskonen, {\it {Phase transition and
  gravitational wave phenomenology of scalar conformal extensions of the
  Standard Model}},  {\em Eur. Phys. J.} {\bf C77} (2017), no.~7 484,
  [\href{http://arxiv.org/abs/1704.01034}{{\tt arXiv:1704.01034}}].

\bibitem{Bezrukov:2007ep}
F.~L. Bezrukov and M.~Shaposhnikov, {\it {The Standard Model Higgs boson as the
  inflaton}},  {\em Phys. Lett.} {\bf B659} (2008) 703--706,
  [\href{http://arxiv.org/abs/0710.3755}{{\tt arXiv:0710.3755}}].

\bibitem{Bezrukov:2010jz}
F.~Bezrukov, A.~Magnin, M.~Shaposhnikov, and S.~Sibiryakov, {\it {Higgs
  inflation: consistency and generalisations}},  {\em JHEP} {\bf 01} (2011)
  016, [\href{http://arxiv.org/abs/1008.5157}{{\tt arXiv:1008.5157}}].

\bibitem{Martin2013a}
J.~Martin, H.~Motohashi, and T.~Suyama, {\it {Ultra Slow-Roll Inflation and the
  non-Gaussianity Consistency Relation}},  {\em Phys. Rev.} {\bf D87} (2013),
  no.~2 023514, [\href{http://arxiv.org/abs/1211.0083}{{\tt arXiv:1211.0083}}].

\bibitem{Motohashi2015a}
H.~Motohashi, A.~A. Starobinsky, and J.~Yokoyama, {\it {Inflation with a
  constant rate of roll}},  {\em JCAP} {\bf 1509} (2015), no.~09 018,
  [\href{http://arxiv.org/abs/1411.5021}{{\tt arXiv:1411.5021}}].

\bibitem{Davydov2016}
E.~Davydov and D.~Gal'tsov, {\it {HYM-flation: Yang–Mills cosmology with
  Horndeski coupling}},  {\em Phys. Lett.} {\bf B753} (2016) 622--628,
  [\href{http://arxiv.org/abs/1512.02164}{{\tt arXiv:1512.02164}}].

\bibitem{Motohashi2017a}
H.~Motohashi and A.~A. Starobinsky, {\it {Constant-roll inflation:
  confrontation with recent observational data}},  {\em Europhys. Lett.} {\bf
  117} (2017), no.~3 39001, [\href{http://arxiv.org/abs/1702.05847}{{\tt
  arXiv:1702.05847}}].

\bibitem{Gao2017}
Q.~Gao and Y.~Gong, {\it {Reconstruction of extended inflationary potentials
  for attractors}},  \href{http://arxiv.org/abs/1703.02220}{{\tt
  arXiv:1703.02220}}.

\bibitem{Odintsov2017}
S.~D. Odintsov and V.~K. Oikonomou, {\it {Inflationary Dynamics with a Smooth
  Slow-Roll to Constant-Roll Era Transition}},  {\em JCAP} {\bf 1704} (2017),
  no.~04 041, [\href{http://arxiv.org/abs/1703.02853}{{\tt arXiv:1703.02853}}].

\bibitem{Odintsov2017a}
S.~D. Odintsov and V.~K. Oikonomou, {\it {Inflation with a Smooth Constant-Roll
  to Constant-Roll Era Transition}},  {\em Phys. Rev.} {\bf D96} (2017), no.~2
  024029, [\href{http://arxiv.org/abs/1704.02931}{{\tt arXiv:1704.02931}}].

\bibitem{Nojiri2017}
S.~Nojiri, S.~D. Odintsov, and V.~K. Oikonomou, {\it {Constant-roll Inflation
  in $F(R)$ Gravity}},  {\em Class. Quant. Grav.} {\bf 34} (2017), no.~24
  245012, [\href{http://arxiv.org/abs/1704.05945}{{\tt arXiv:1704.05945}}].

\bibitem{Motohashi2017b}
H.~Motohashi and A.~A. Starobinsky, {\it {$f(R)$ constant-roll inflation}},
  {\em Eur. Phys. J.} {\bf C77} (2017), no.~8 538,
  [\href{http://arxiv.org/abs/1704.08188}{{\tt arXiv:1704.08188}}].

\bibitem{Gao2017a}
Q.~Gao, {\it {Reconstruction of constant slow-roll inflation}},  {\em Sci.
  China Phys. Mech. Astron.} {\bf 60} (2017), no.~9 090411,
  [\href{http://arxiv.org/abs/1704.08559}{{\tt arXiv:1704.08559}}].

\bibitem{Oikonomou2017}
V.~K. Oikonomou, {\it {Reheating in Constant-roll $F(R)$ Gravity}},  {\em Mod.
  Phys. Lett.} {\bf A32} (2017), no.~33 1750172,
  [\href{http://arxiv.org/abs/1706.00507}{{\tt arXiv:1706.00507}}].

\bibitem{Odintsov2017b}
S.~D. Odintsov, V.~K. Oikonomou, and L.~Sebastiani, {\it {Unification of
  Constant-roll Inflation and Dark Energy with Logarithmic $R^2$-corrected and
  Exponential $F(R)$ Gravity}},  {\em Nucl. Phys.} {\bf B923} (2017) 608--632,
  [\href{http://arxiv.org/abs/1708.08346}{{\tt arXiv:1708.08346}}].

\bibitem{Oikonomou2017a}
V.~K. Oikonomou, {\it {A Smooth Constant-Roll to a Slow-Roll Modular Inflation
  Transition}},  {\em Int. J. Mod. Phys.} {\bf D27} (2017), no.~02 1850009,
  [\href{http://arxiv.org/abs/1709.02986}{{\tt arXiv:1709.02986}}].

\bibitem{Cicciarella2018}
F.~Cicciarella, J.~Mabillard, and M.~Pieroni, {\it {New perspectives on
  constant-roll inflation}},  {\em JCAP} {\bf 1801} (2018), no.~01 024,
  [\href{http://arxiv.org/abs/1709.03527}{{\tt arXiv:1709.03527}}].

\bibitem{Awad2017}
A.~Awad, W.~El~Hanafy, G.~G.~L. Nashed, S.~D. Odintsov, and V.~K. Oikonomou,
  {\it {Constant-roll Inflation in $f(T)$ Teleparallel Gravity}},
  \href{http://arxiv.org/abs/1710.00682}{{\tt arXiv:1710.00682}}.

\bibitem{Anguelova2017}
L.~Anguelova, P.~Suranyi, and L.~C.~R. Wijewardhana, {\it {Systematics of
  Constant Roll Inflation}},  {\em JCAP} {\bf 1802} (2017), no.~02 004,
  [\href{http://arxiv.org/abs/1710.06989}{{\tt arXiv:1710.06989}}].

\bibitem{Ito2017}
A.~Ito and J.~Soda, {\it {Anisotropic Constant-roll Inflation}},  {\em Eur.
  Phys. J.} {\bf C78} (2017), no.~1 55,
  [\href{http://arxiv.org/abs/1710.09701}{{\tt arXiv:1710.09701}}].

\bibitem{Yi2017}
Z.~Yi and Y.~Gong, {\it {On the constant-roll inflation}},
  \href{http://arxiv.org/abs/1712.07478}{{\tt arXiv:1712.07478}}.

\bibitem{Mohammadi2018}
A.~Mohammadi, K.~Saaidi, and T.~Golanbari, {\it {Tachyon constant-roll
  inflation}},  \href{http://arxiv.org/abs/1801.03487}{{\tt arXiv:1801.03487}}.

\bibitem{Gao2018a}
Q.~Gao, Y.~Gong, and Q.~Fei, {\it {Constant-roll tachyon inflation and
  observational constraints}},  \href{http://arxiv.org/abs/1801.09208}{{\tt
  arXiv:1801.09208}}.

\bibitem{Gao2018}
Q.~Gao, {\it {The observational constraint on constant-roll inflation}},
  \href{http://arxiv.org/abs/1802.01986}{{\tt arXiv:1802.01986}}.

\bibitem{Anguelova2016}
L.~Anguelova, {\it {A Gravity Dual of Ultra-slow Roll Inflation}},  {\em Nucl.
  Phys.} {\bf B911} (2016) 480--499,
  [\href{http://arxiv.org/abs/1512.08556}{{\tt arXiv:1512.08556}}].

\bibitem{Anguelova2016a}
L.~Anguelova, {\it {Glueball Inflation and Gauge/Gravity Duality}},  {\em
  Springer Proc. Math. Stat.} {\bf 191} (2016) 285--293,
  [\href{http://arxiv.org/abs/1601.02449}{{\tt arXiv:1601.02449}}].

\bibitem{Cai2016}
Y.-F. Cai, J.-O. Gong, D.-G. Wang, and Z.~Wang, {\it {Features from the
  non-attractor beginning of inflation}},  {\em JCAP} {\bf 1610} (2016), no.~10
  017, [\href{http://arxiv.org/abs/1607.07872}{{\tt arXiv:1607.07872}}].

\bibitem{Gong2017}
Y.~Gong, {\it {Primordial black holes from ultra-slow-roll inflation}},
  \href{http://arxiv.org/abs/1707.09578}{{\tt arXiv:1707.09578}}.

\bibitem{Hirano2016}
S.~Hirano, T.~Kobayashi, and S.~Yokoyama, {\it {Ultra slow-roll G-inflation}},
  {\em Phys. Rev.} {\bf D94} (2016), no.~10 103515,
  [\href{http://arxiv.org/abs/1604.00141}{{\tt arXiv:1604.00141}}].

\bibitem{Tsamis2004}
N.~C. Tsamis and R.~P. Woodard, {\it {Improved estimates of cosmological
  perturbations}},  {\em Phys. Rev.} {\bf D69} (2004) 084005,
  [\href{http://arxiv.org/abs/astro-ph/0307463}{{\tt astro-ph/0307463}}].

\bibitem{Liddle1994}
A.~R. Liddle, P.~Parsons, and J.~D. Barrow, {\it {Formalizing the slow roll
  approximation in inflation}},  {\em Phys. Rev.} {\bf D50} (1994) 7222--7232,
  [\href{http://arxiv.org/abs/astro-ph/9408015}{{\tt astro-ph/9408015}}].

\bibitem{Jaerv2015}
L.~J{\"a}rv, P.~Kuusk, M.~Saal, and O.~Vilson, {\it Invariant quantities in the
  scalar-tensor theories of gravitation},  {\em Physical Review D} {\bf 91}
  (2015), no.~2 024041, [\href{http://arxiv.org/abs/1411.1947}{{\tt
  arXiv:1411.1947}}].

\bibitem{Kuusk2016}
P.~Kuusk, M.~R{\"u}nkla, M.~Saal, and O.~Vilson, {\it {Invariant slow-roll
  parameters in scalar--tensor theories}},  {\em Class. Quant. Grav.} {\bf 33}
  (2016), no.~19 195008, [\href{http://arxiv.org/abs/1605.07033}{{\tt
  arXiv:1605.07033}}].

\bibitem{Kuusk2016a}
P.~Kuusk, L.~Jarv, and O.~Vilson, {\it {Invariant quantities in the
  multiscalar-tensor theories of gravitation}},  {\em Int. J. Mod. Phys.} {\bf
  A31} (2016), no.~02n03 1641003, [\href{http://arxiv.org/abs/1509.02903}{{\tt
  arXiv:1509.02903}}].

\bibitem{Jaerv2017}
L.~J{\"a}rv, K.~Kannike, L.~Marzola, A.~Racioppi, M.~Raidal, M.~R{\"u}nkla,
  M.~Saal, and H.~Veerm{\"a}e, {\it {Frame-Independent Classification of
  Single-Field Inflationary Models}},  {\em Phys. Rev. Lett.} {\bf 118} (2017),
  no.~15 151302, [\href{http://arxiv.org/abs/1612.06863}{{\tt
  arXiv:1612.06863}}].

\bibitem{Karam2017}
A.~Karam, T.~Pappas, and K.~Tamvakis, {\it {Frame-dependence of higher-order
  inflationary observables in scalar-tensor theories}},  {\em Phys. Rev.} {\bf
  D96} (2017), no.~6 064036, [\href{http://arxiv.org/abs/1707.00984}{{\tt
  arXiv:1707.00984}}].

\bibitem{Remazeilles:2017szm}
{\bf CORE} Collaboration, M.~Remazeilles et~al., {\it {Exploring Cosmic Origins
  with CORE: B-mode Component Separation}},
  \href{http://arxiv.org/abs/1704.04501}{{\tt arXiv:1704.04501}}.

\end{thebibliography}\endgroup

\end{document}